\documentclass{elsart}
\usepackage{graphicx}
\usepackage{amssymb}

\begin{document}

\begin{frontmatter}

\title{Universality Class of Bak-Sneppen Model on Scale-Free Network}

\author[jwl]{Kyoung Eun Lee}
\author[hong]{Byoung Hee Hong}
\author[jwl,par1]{Jae Woo Lee\corauthref{cor1}}
\ead{jaewlee@inha.ac.kr}

\corauth[cor1]{Corresponding author. Tel.: 82+32+8607660; Fax: 82+32+8727562}
\address[jwl]{Department of Physics, Inha University, Incheon 402-751, Korea}
\address[hong]{Institute of Technology Woo Young Co. LTD, KAIST Academica-Industry 
Research Hall, Seoul 130-150, Korea}
\address[par1]{School of Computational Science,  
Florida State University, Tallahassee FL 32306-4120, USA}

\begin{abstract}

  We study the critical properties of the Bak-Sneppen coevolution model on scale-free
networks by Monte Carlo method. We report the distribution of the avalanche 
size and fractal activity through the branching process. We observe that the
critical fitness $f_c (N)$ depends on the number of the node such as
$f_c (N) \sim 1/ \log (N)$ for both the scale-free network and the
directed scale-free network. Near the critical fitness many physical quantities 
show power-law behaviors. 
The probability distribution $P(s)$ of the avalanche 
size at the critical fitness shows a 
power-law like $P(s) \sim s^{-\tau}$ with $\tau=1.80(3)$ regardless of the 
scale-free network and the directed scale free network.
The probability distribution $P_f (t)$ of
the first return time also shows a power-law such as
$P_f (t) \sim t^{-\tau_f}$.  The probability distribution of the first return time
has two scaling regimes. The critical exponents $\tau_f$ are
equivalent for both the scale-free network and the directed scale-free network.
We obtain the critical exponents as $\tau_{f1} =2.7(1)$ at $t < t_c$
and $\tau_{f2} = 1.72(3)$ at $ t >t_c$ where the crossover time
$t_c \sim 100$. 
The Bak-Sneppen model on the scale-free network and directed scale-free network
shows a unique universality class. The critical exponents are different from
the mean-field results. The directionality of the network does not change the
universality on the network. 
\end{abstract}

\end{frontmatter}

\newcommand{\be}{\begin{equation}}
\newcommand{\ee}{\end{equation}}

\section{Introduction}

 The concepts of self-organized criticality (SOC) has been widely applied
in nonequilibrium systems such as biological evolution, slowly driven 
systems and economic 
systems\cite{Bak99,JE98,PMB96,MA99,MA97,BP00,Lee1,Lee2}.
 In the SOC there is no characteristic
scale of the system. The physical quantities show the power-law behaviors.
The Bak-Sneppen (BS) model is a typical model showing 
the SOC\cite{BTW87,BS93}. 
The BS model is a simple coevolution model of the biological species evolving
with fitness. The avalanche size distribution $P(s)$ is characterized
by a power-law like $P(s) \sim s^{-\tau}$ where $\tau$ is an avalanche
critical exponent\cite{PMB96,BS93}. 
The critical exponent $\tau$ depends on the 
dimensionality of the lattice. In the lattice BS model the directionality
of the update of the fitness influences the critical exponents of the
avalanche size distribution. The critical exponents of the fully anisotropic
BS model are different to those of the isotropic BS model\cite{TL04,GD04}.

  Recently, complex networks have been found in many systems such as 
WWW, social networks, biological networks,
 etc.\cite{WS98,BA99,AB02,DM02,NE03}. 
Such networks is called scale-free network (SFN).  The degree
distribution of the SFN follow the power-law like 
$P(k) \sim k^{-\gamma} $ with $\gamma > 2$ where $k$ is a 
degree of the node on the SFN\cite{BA99,AB02,DM02,NE03}.
In the SFN there are
hubs with large degrees. The degree distribution and the existence of the hubs
contribute to the dynamics on the SFN. 
Moreno and Vazquez reported the critical fitness and the critical exponent $\tau$
of the BS model on Barabasi-Albert (BA) network. They observed that the critical exponent
$\tau$ is close to the mean-field value $3/2$\cite{MV02}.
Lee and Kim also reported the critical fitness $f_c (N) \sim 1/ \langle (k+1)^2 \rangle $
on the SFN with the size $N$ for $2 < \gamma \le 3$. They observed
two power-law regimes of the avalanche size distribution\cite{LK05}.
Masuda et al. obtained the analytic solution of the critical 
fitness $f_c \sim 1/ ( \langle k^2 \rangle / \langle k \rangle +1)$ and
$\tau=3/2$\cite{MGK05}.
In this article we study the BS model on the SFN and DSFN
by Monte Carlo method. 
We observe that the critical fitness shows the logarithmic dependence of
the total number of the node on the SFN. 
The avalanche size distribution follows the power-law. The
critical exponent $\tau$ is independent of the directionality of the SFN. 
We observe the two scaling regimes in the probability distribution 
of the first return time. We observe that the BS model on the SFN and DSFN belongs
to the same universality class. 
In section 2 we introduce the BS model and SFN. In section 2 we report the critical
exponents of the BS model on the SFN and DSFN. We give the concluding remarks
in section 4.

\section{Bak-Sneppen model on scale-free network}

Consider a scale-free network with a degree distribution $p(k) \sim k^{-\gamma}$ with a
exponent $\gamma=3.0$. The SFN has the different geometrical properties in comparison
to the regular lattice, small-world network, and random network.
The SFN shows the scale-free behavior of the node degree distribution and small-world 
behaviors. We consider  a scale-fee network and a directed scale-free network.
We construct the SFN by Barabasi-Albert algorithm: (i) put $m_o = 3$ 
initial nodes. (ii) put a new node of $m=3$ links with connecting 
probability $p_i =k_i / \sum_j k_j $ where $k_i$ is the degree of the node
(the number of links connecting to node $i$). (iii) repeat step 2
until the total number of nodes are $N=10^5$. 
The directed SFN is obtained by applying a random direction of each links of 
the SFN. 

\begin{figure}
\includegraphics[width=10cm,height=12cm,angle=270,clip]{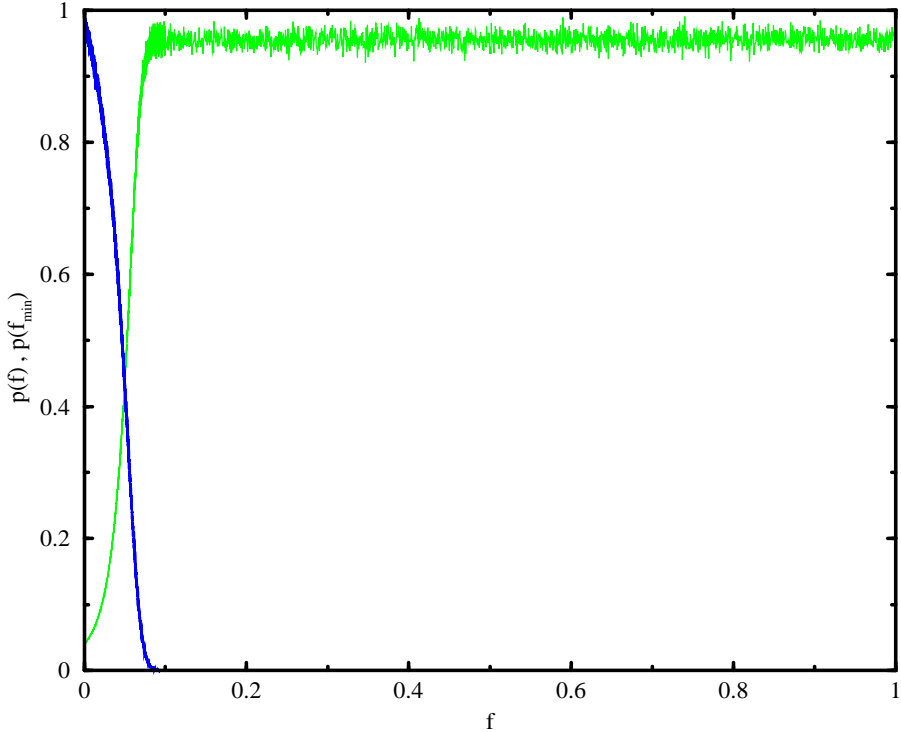}
\caption[0]{
The probability density of the fitness $p(f)$ and the least fitness $p(f_{\rm min})$
on the SFN with the total number of nodes $N=512$ for the Bak-Sneppen model. We obtain the 
critical fitness $f_c = 0.10413(5)$.
}
\label{fig1}
\end{figure}

\begin{figure}
\includegraphics[width=10cm,height=12cm,angle=270,clip]{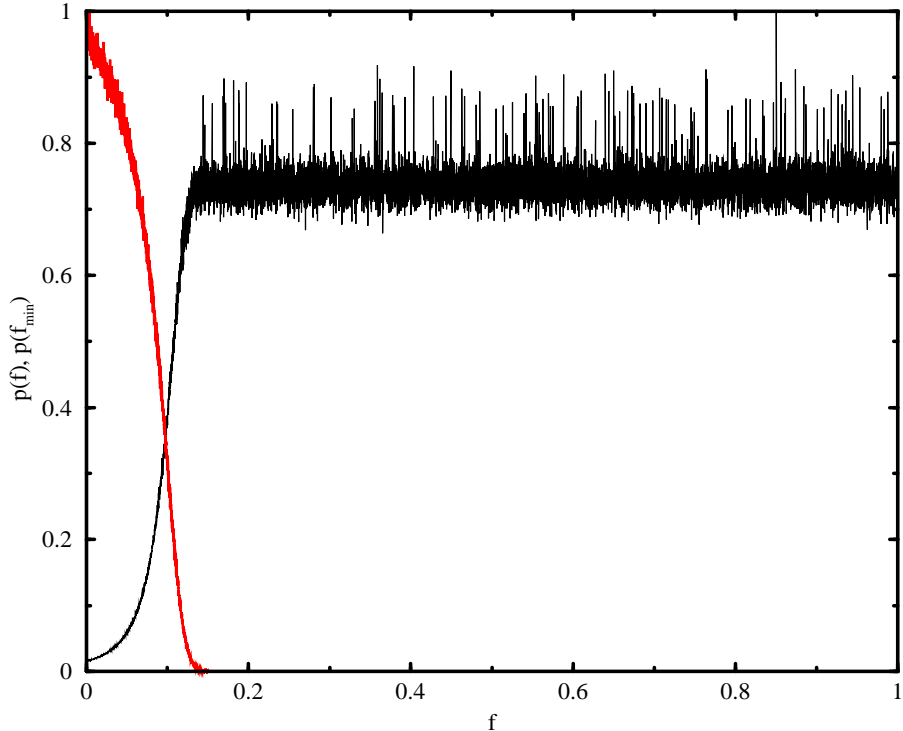}
\caption[0]{
The probability density of the fitness $p(f)$ and the least fitness $p(f_{\rm min})$
on the DSFN with the total number of nodes $N=512$ for the  Bak-Sneppen model.
We obtain the critical fitness $f_c = 0.1349(5)$.
}
\label{fig2}
\end{figure}

 We apply the BS coevolution model on the networks.
The BS model is a simple coevolution model which shows punctuated 
equilibrium and the self-organized criticality.
Let's assign random fitness $f_i$ (a uniform real random number 
$0 < f_i < 1$) on the node of the network. Search a node with the least fitness.
Update the fitness of the node with the least fitness and its nearest neighbor nodes.
Repeat these steps. The system goes to a steady-state.  At the steady-state
the probability density that the system lies $f < f_c $ vanishes and 
is uniform above $f_c$ where $f_c$ is a critical fitness. 
We show the probability density $P(f)$ of the fitness  and the probability
density $P(f_{\rm min})$  of the least fitness  with the total number of nodes $N=512$ 
on the SFN in Fig. \ref{fig1} and DSFN in Fig. \ref{fig2}.
The probability density $P(f)$ of the fitness is uniform at $f>f_c$.
The fluctuations of the probability
density $P(f)$ are very large at $f> f_c$ on the DSFN because of the 
directed connections of the links.
The probability density $P(f_{\rm min})$ of the least fitness drops to 
zero at the critical fitness.

\begin{figure}
\includegraphics[width=10cm,height=12cm,angle=270,clip]{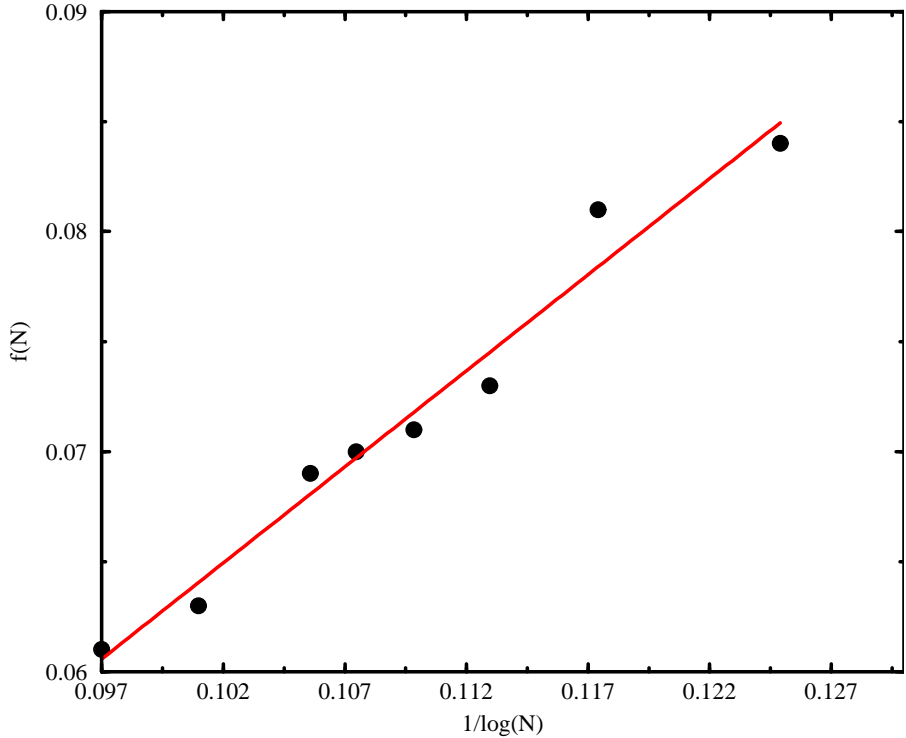}
\caption[0]{
The critical fitness vs the number of the nodes on the network. 
We observe $f_c \sim 1 / \log (N)$. At $N \rightarrow \infty$
$f_c$ goes to zero.
}
\label{fig3}
\end{figure}

We obtained the critical fitness as $f_c =0.10413(5)$ on the SFN and 
$f_c =0.1349(5)$ on the DSFN for $N=512$. In the SFN the critical fitness depends on the 
total number of nodes $N$. In the thermodynamic limit $N \rightarrow \infty$,
the critical fitness approaches logarithmically to zero. 
In Fig. \ref{fig3} we present the critical fitness as a function of the number
of the nodes $N$. We observed $f_c \sim 1/ \ln N$ and $f_c =0$ at 
$N \rightarrow \infty$ for both the SFN and DSFN.
This result for the SFN is consistent with the observations of Mareno and Vazquez\cite{MV02}.
Masuda et al. derived the critical fitness of isotropic BS model 
such as $f_c \sim 1/ (<k^2 >/<k> +1)$ for the SFN. 
In the finite number of node $N$, the average number of nodes for the SFN is given
as $<k^2 > \sim \ln N$ and $<k> \sim {\rm constant}$. Our result of 
the critical fitness is consistent with the predictions of Masuda et al.\cite{MGK05}.
The critical fitness for the DSFN also follows the logarithmic dependence.
 So, we observe the critical 
behaviors at the finite critical point $f_c (N)$ in the simulation.

\section{Critical Exponents and Universality Class}
 
 In the BS model the branching processes simplify the dynamic evolutions. Consider
an $f_o$-avalanche. In $f_o$-avalanche all the sites with $f_i > f_o$ are treated
as vacuum sites. The sites with $f_i < f_o$ are active sites. In the branching process
we first create a random number, chosen from the uniform distribution, at a randomly 
selected node on the network. All other sites are vacuum nodes. At each
time step, the site of the minimal fitness with $f_i < f_o$ activates 
until there are no more nodes with $f_i < f_o$. If there is no an active site anymore, 
the $f_o$-avalanche is finished. 

\begin{figure}
\includegraphics[width=10cm,height=12cm,angle=270,clip]{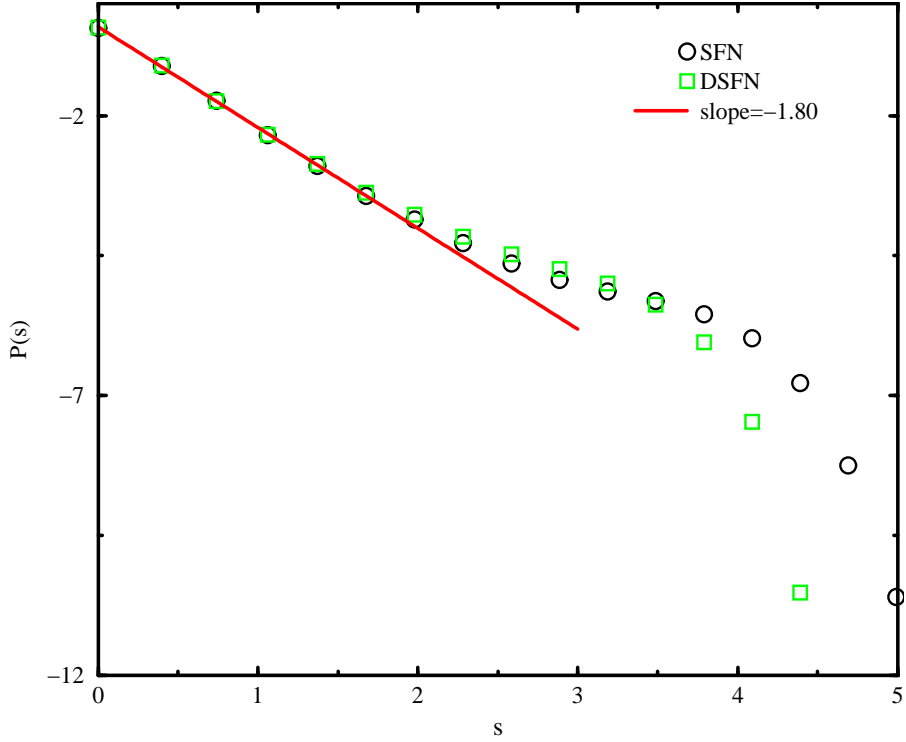}
\caption[0]{
The log-log plot of the probability distribution function $P(s)$
of the avalanche size  as a function of
the avalanche size $s$ at the critical fitness on the SFN and DSFN.
}
\label{fig4}
\end{figure}

 We measure the distribution of the avalanche size by the branching process. 
In  the lattice BS model the probability distribution 
$P(s, f_o )$ of
the $f_o$-avalanche of the size $s$ is described by the scaling function\cite{PMB96}
\be
P(s,f_o ) = s^{-\tau} g(s(f_c - f_o )^{1/\sigma} )
\ee
where $\tau$ and $\sigma$ are critical exponents and $g(x)$ is a 
scaling function. 
The scaling function is given by $g(x) \rightarrow 0$ for $x >>  1$
and $g(x) \rightarrow {\rm constant}$ for $x \rightarrow 0$. 
At the critical fitness $f_o \rightarrow f_c$, the probability distribution
of the avalanche follows the power-law as
\be
P(s) \sim s^{-\tau}
\ee
In Fig. \ref{fig4} we present the probability distribution $P(s)$ of
the avalanche size $s$ at the critical fitness on the SFN and DSFN. 
The probability distribution 
$P(s)$ follows the power-law like $P(s) \sim s^{-\tau}$ up to 
the avalanche size $s_c =100$. 
We plot the histogram of the avalanche size distribution
by exponentially increasing bins to reduce the fluctuations of the
data\cite{RK}.
We obtain the critical
exponent $\tau=1.80(3)$ both the SFN and DSFN. 
At the large avalanche size the probability distribution $P(s)$
decays according to the exponential function.
The observed critical exponent $\tau$ is different from
the mean-field value $\tau=3/2$ and bigger than the exponents
of the lattice BS model. 
The scale-free network contributes the enhancement of the 
critical exponent $\tau$. 
In our results we observe only a scaling regime at the range
$ 1<s<100$. At the large $s$ the avalanche size distribution deviates
from the power-law and follow exponential decays. We can not
observe two scaling regimes reported by Lee and Kim\cite{LK05}.
The critical exponents $\tau$ are the same for both the SFN and DSFN.
The directionality of the network does not change the critical behavior
of the probability distribution of the avalanche size. 

\begin{figure}
\includegraphics[width=10cm,height=12cm,angle=270,clip]{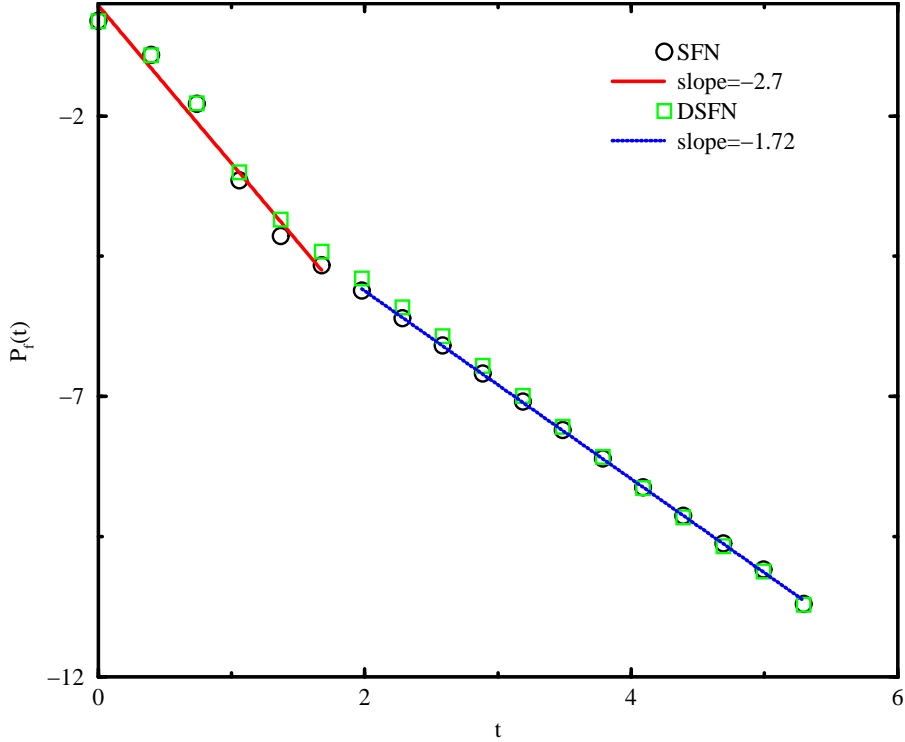}
\caption[0]{
The log-log plot of the probability distribution function $P(s)$
of the first return time  as a function of
the time $t$ at the critical fitness on SFN and DSFN.
}
\label{fig5}
\end{figure}

In Fig. \ref{fig5} we present the probability distribution of the 
first return time. Consider an active site at a time $t^{\prime}$.
This site is active again at a time $t^{\prime}+t$. We define
the time interval $t$ as the first return time (or waiting time). 
In the random process the probability distribution of the first
return time is Poisson distribution. However, in the SOC system
the probability distribution $P_f (t)$ of the first return time
follows a power-law like
\be
P_f (t) \sim t^{-\tau_f}
\ee
where $\tau_f$ is a critical exponent. 
In Fig. \ref{fig5} we can observe the obvious two scaling 
regimes with the crossover time $t_c \sim 100$ for both the SFN and DSFN.
We obtain the critical exponent $\tau_{f1} = 2.7(1)$ at the early-time
regime and $\tau_{f2} = 1.72(3)$ at the late-time regime for 
both the SFN and DSFN.
The critical exponent $\tau_{f2}$ is greater than the exponent
$\tau_f =1.58$ on one dimensional lattice and $\tau_f = 1.28$
on two dimensional lattice.

The observed critical exponents on the SFN and DSFN are different
from the mean-field predictions and those of the lattice
BS model. We conclude that the BS model on the SFN and DSFN belongs
to a unique universality class. The directionality of the network
does not change the critical exponents. 

\section{Conclusion}

We consider the Bak-Sneppen coevolution model on the SFN and DSFN. We obtain
the critical fitness, the probability distribution of the avalanche size,
and the probability distribution of the first return time by the
Monte Carlo simulations.
We observe that the critical fitness depend on the total number of the
nodes for both the SFN and DSFN. The critical fitness goes to zero according to
$f_c \sim 1/ \log(N)$. 
The avalanche size distribution decays according to the power-law
at the early scaling regime. The critical exponent of avalanche size
is greater than that of the lattice BS model.
The probability distribution of the first return time
has two scaling regimes separated by the crossover time $t_c \sim 100$.
We have found that the Bak-Sneppen model on the SFN and DSFN belongs
to a new universality class. The directionality of the networks
does not change the universality for the Bak-Sneppen model on the SFN.
 
\section*{Acknowledgments}
The present work has been supported by a research grant of the Asan
Foundation.

\newcommand{\jpa}{J. Phys. A}
\newcommand{\jkps}{J. Kor. Phys. Soc.}


\begin{thebibliography}{}
\bibitem{Bak99} P. Bak, How nature works: the science of self-organized criticality, 
Springer-Verlag, New York, 1999.
\bibitem{JE98} H. J. Jensen, Self-organized criticality: Emergent complex behavior
in physical and biological systems, Cambridge University Press, Cambridge, 1998.
\bibitem{PMB96} M. Paczuski, S. Maslov, P. Bak, Phys. Rev. E 53 (1996) 414.
\bibitem{MA99} R. N. Mantegna and H. E. Stanley,  An Introduction to
Econophysics: Correlations and Complexity in Finance, Cambridge University Press,
Cambridge, 1999.
 \bibitem{MA97} B. Mandelbrot, Fractals and Scaling in Finance, Springer,
New York, 1997.
\bibitem{BP00} J. P. Bouchaud and M. Potters, Theory of Financial
Risk, Cambridge University Press, New York, 2000.
\bibitem{Lee1} K. E. Lee, J. W. Lee, J. Kor. Phys. Soc. 46 (2005) 726.
\bibitem{Lee2} K. E. Lee, J. W. Lee, J. Kor. Phys. Soc. 47 (2005) 185.
\bibitem{BTW87} P. Bak, C. Tang, K. Wiesenfeld, Phys. Rev. Lett. 59 (1987) 381.
\bibitem{BS93} P. Bak, K. Sneppen, Phys. Rev. Lett. 71 (1993) 4083.
\bibitem{TL04} U. Tirnakli, M. L. Lyra, Physica A 342 (2004) 151.
\bibitem{GD04} G. J. M. Garcia, R. Dickman, Physica A 342 (2004) 516.
\bibitem{WS98} D. J. Watts, S. H. Strogatz, Nature 393 (1998) 440.
\bibitem{BA99} A.-L. Barab\'{a}si, R. Albert, Science 286 (1999) 509.
\bibitem{AB02} R. Albert, Barab\'{a}si, Rev. Mod. Phys. 74 (2002) 47.
\bibitem{DM02} S. N. Dorogovtsev, J. J. F. Mendes, Adv. Phys. 51 (2002) 1079.
\bibitem{NE03} M. E. J. Newman, SIAM Rev. 45 (2003) 167.
\bibitem{MV02} Y. Moreno, A. Vazquez, Europhys. Lett. 57 (2002) 765.
\bibitem{LK05} S. Lee, Y. Kim, Phys. Rev. E 71 (2005) 057102.
\bibitem{MGK05} N. Masuda, K.-I. Goh, B. Kahng, cond-mat/0508623.
\bibitem{RK} P. A. Rikvold, R. K. P. Zia, Phys. Rev. E 68 (2003) 031013.
\end{thebibliography}
\end{document}